\documentclass{article}
\usepackage{amsmath}
\usepackage{appendix}
\usepackage[retainorgcmds]{IEEEtrantools}
\usepackage{graphicx}
\begin{document}
\title{Exact canonic eigenstates of the truncated Bogoliubov Hamiltonian in an interacting bosons gas}
\author{Loris Ferrari \\ Department of Physics and Astronomy of the University (DIFA) \\Via Irnerio, 46,40127, Bologna, Italy}
\maketitle
\begin{abstract}
In a gas of $N$ weakly interacting bosons \cite{Bogo1, Bogo2}, a truncated canonic Hamiltonian $\widetilde{h}_c$ follows from dropping all the interaction terms between free bosons with momentum $\hbar\mathbf{k}\ne\mathbf{0}$. Bogoliubov Canonic Approximation (BCA) is a further manipulation, replacing the number \emph{operator} $\widetilde{N}_{in}$ of free particles in $\mathbf{k}=\mathbf{0}$, with the total number $N$ of bosons. BCA transforms $\widetilde{h}_c$ into a different Hamiltonian $H_{BCA}=\sum_{\mathbf{k}\ne\mathbf{0}}\epsilon(k)B^\dagger_\mathbf{k}B_\mathbf{k}+const$, where $B^\dagger_\mathbf{k}$ and $B_\mathbf{k}$ create/annihilate non interacting pseudoparticles. The problem of the \emph{exact} eigenstates  of the truncated Hamiltonian is completely solved in the thermodynamic limit (TL) for a special class of eigensolutions $|\:S,\:\mathbf{k}\:\rangle_{c}$, denoted as \textquoteleft s-pseudobosons\textquoteright, with energies $\mathcal{E}_{S}(k)$ and \emph{zero} total momentum. Some preliminary results are given for the exact eigenstates (denoted as \textquoteleft $\eta$-pseudobosons\textquoteright), carrying a total momentum $\eta\hbar\mathbf{k}$ ($\eta=\:1,\:2,\: \dots$). A comparison is done with $H_{BCA}$ and with the Gross-Pitaevskii theory (GPT), showing that some differences between exact and BCA/GPT results persist even in the TL. Finally, it is argued that the emission of $\eta$-pseudobosons, which is responsible for the dissipation $\acute{a}$ \emph{la} Landau~\cite{L}, could be significantly different from the usual picture, based on BCA pseudobosons.\newline
\end{abstract}

e-mail: loris.ferrari@unibo.it
telephone: ++39-051-2095109

\section{Introduction}
\label{intro}

The Hamiltonian of a gas of $N$ interacting bosons of mass $M$ in a volume $V$ reads: 

\begin{equation}
H_{bos}=\sum_{\mathbf{k}}\overbrace{\left(\hbar^2k^2/2M\right)}^{\mathcal{T}(k)}b^\dagger_{\mathbf{k}}b_{\mathbf{k}}+\frac{1}{2}\sum_{\mathbf{k}_1,\mathbf{k}_2,\mathbf{q}}\widehat{u}(q)\:b^\dagger_{\mathbf{k}_2-\mathbf{q}}b^\dagger_{\mathbf{k}_1+\mathbf{q}}b_{\mathbf{k}_1}b_{\mathbf{k}_2}\:,\label{Hbos}
\end{equation}
\\
where $b^\dagger_{\mathbf{k}}$ and $b_{\mathbf{k}}$ create and destroy a spinless boson in the free-particle state $\langle\:\mathbf{r}\:|\:\mathbf{k}\:\rangle= e^{i\mathbf{k}\:\mathbf{r}}/\sqrt{V}$
and  

\begin{equation*}
\widehat{u}(q)=\frac{1}{V}\int\mathrm{d}\mathbf{r}e^{-i\mathbf{q}\:\mathbf{r}}\:u(r)\:,
\end{equation*}
\\
is the Fourier transform of the \emph{repulsive} interaction energy $u(r)$ ($>0)$.\newline

Bogoliubov's approach to the study of Hamiltonian~\eqref{Hbos} results in dropping all the interaction terms that couple bosons in the excited states  $|\:\mathbf{k}\:\rangle$, which we call the First Bogoliubov's Approximation (FBA). This leads to the Hamiltonian \footnote{Here and in what follows all overtilded symbols $\widetilde{\dots}$ indicate operators.}:                                                                                                                                                                                                                                                                                                                                                                                                                                                                                                                                                                                                                                                                                                                                                                                                                                                                                                                                                                                                                                                                                                                                                                                                                                                                                                                                                                                                                                                                                        

\begin{align}
H_{FBA}&=\frac{\widehat{u}(0)}{2}\left[\widetilde{N}^2-\widetilde{N}_{out}^2\right]\left[1+(\widetilde{N}+\widetilde{N}_{out})^{-1}\right]+\nonumber\\
\nonumber\\
&+\sum_{\mathbf{k}\ne0}\underbrace{\left[\mathcal{T}(k)+\widetilde{N}_{in}\:\widehat{u}(k)\right]}_{\widetilde{\epsilon}_1(k)}b^\dagger_{\mathbf{k}}b_{\mathbf{k}}+\nonumber\\
\nonumber\\
&+\frac{1}{2}\sum_{\mathbf{k}\ne0}\widehat{u}(k)\Big[b^\dagger_{\mathbf{k}}b^\dagger_{-\mathbf{k}}(\:b_{\mathbf{0}}\:)^2+b_{\mathbf{k}}b_{-\mathbf{k}}(\:b^\dagger_{\mathbf{0}}\:)^2\Big]\:,\label{HFBA}
\end{align}
\\
where 

\begin{equation}
\widetilde{N}_{in}=b^\dagger_{\mathbf{0}}b_{\mathbf{0}}\quad;\quad\widetilde{N}_{out}=\sum_{\mathbf{k}\ne0}b^\dagger_{\mathbf{k}}b_{\mathbf{k}}
\end{equation}
\\
are number operators and $\widetilde{N}=\widetilde{N}_{in}+\widetilde{N}_{out}$ is the total number of bosons, with conserved value $N$, in the canonic case.

The Hamiltonian $H_{FBA}$ is the starting point of a split treatment of the interacting bosonic gas, under \emph{canonic} or \emph{grand canonic} conditions \cite{AB}. The main difference is that in the canonic case the operator $\widetilde{N}$ is (rigorously) replaced by a conserved c-number $N$, while in the grand canonic case the operators replaced by c-numbers ($C$, $C^*$)  are $(\:b_{\mathbf{0}}\:)^2$ and $(\:b^\dagger_{\mathbf{0}}\:)^2$. This is what we call the Second Bogoliubov Approximation (SBA), that yields the grand canonic Hamiltonian:

\begin{align}
H_{SBA}&=\frac{\widehat{u}(0)}{2}\left[\widetilde{N}^2-\widetilde{N}_{out}^2\right]\left[1+(\widetilde{N}+\widetilde{N}_{out})^{-1}\right]+\nonumber\\
\nonumber\\
&+\sum_{\mathbf{k}\ne0}\widetilde{\epsilon}_1(k)\:b^\dagger_{\mathbf{k}}b_{\mathbf{k}}+\frac{1}{2}\sum_{\mathbf{k}\ne0}\widehat{u}(k)\Big[C\:b^\dagger_{\mathbf{k}}b^\dagger_{-\mathbf{k}}+C^*b_{\mathbf{k}}b_{-\mathbf{k}}\Big]\:,\label{HSBA}
\end{align}
\\
where the number of particles is not conserved, and a chemical potential $\mu$ is to be included, as an additional parameter. 

Given the Fock space, spanned by states $|\:N_{in},\:N_{out}\rangle$ with $N_{in}$ bosons in the free-particle ground state $|\:\mathbf{0}\:\rangle$, and $N_{out}=\sum_{\mathbf{k}\ne\mathbf{0}}n_\mathbf{k}$ bosons in the excited states $|\:\mathbf{k}\:\rangle$, the passage from eq.n~\eqref{Hbos} to \eqref{HFBA} results in a low temperature, weak interaction approximation, in which the main sector of Fock space, explored by the gas, is formed by states with $N_{in}\gg N_{out}$. The ratio $\alpha=N_{out}/N_{in}$ is thereby a relevant smallness parameter, that keeps under control a specific order of approximation. For instance, the main condition leading to eq.n~\eqref{HFBA} is that the probability of interaction ($\propto \alpha^2$) between free particles in the excited states is negligible. Hence, the passage from eq.n~\eqref{Hbos} to \eqref{HFBA} is \emph{first-order} in $\alpha$. In the canonic case, this leads one to drop the term $\widetilde{N}_{out}^2/N^2$, for self consistency, and the term $(N+\widetilde{N}_{out})^{-1}$ in the TL ($N\rightarrow\infty$), which transforms eq.n~\eqref{HFBA} into the \emph{canonic} Hamiltonian:

\begin{align}
H_c&=\overbrace{\frac{\widehat{u}(0)N^2}{2}}^{E_{in}}+\sum_{\mathbf{k}\ne0}\overbrace{\left[\mathcal{T}(k)+\widetilde{N}_{in}\:\widehat{u}(k)\right]}^{\widetilde{\epsilon}_1(k)}b^\dagger_{\mathbf{k}}b_{\mathbf{k}}+\nonumber\\
\nonumber\\
&+\frac{1}{2}\sum_{\mathbf{k}\ne0}\widehat{u}(k)\Big[b^\dagger_{\mathbf{k}}b^\dagger_{-\mathbf{k}}(\:b_{\mathbf{0}}\:)^2+b_{\mathbf{k}}b_{-\mathbf{k}}(\:b^\dagger_{\mathbf{0}}\:)^2\Big]\:.\label{Hc}
\end{align}
\\
In ref.~\cite{AB} is shown that a suitable procedure makes the Hamiltonian $H_{c}$ take a non interacting form 

\begin{equation}
H_{BCA}= E_{in}+\sum_{\mathbf{k}\ne0}\overbrace{\left[\epsilon(k)\left(B^\dagger_\mathbf{k}B_\mathbf{k}+\frac{1}{2}\right)-\frac{\epsilon_1(k)}{2}\right]}^{\widetilde{h}_{BCA}(\mathbf{k})}
\:,\label{HBCA''}
\end{equation}
\\
that is interpreted as due to massless \emph{pseudobosons}, created and destroyed by the bosonic operators $B^\dagger_{\mathbf{k}}$,  $B_{\mathbf{k}}$ entering equation \eqref{HBCA''}. It is important to stress that $H_c$ is the \emph{exact} canonic version of $H_{FBA}$ (eq.n~\eqref{HFBA}), while $H_{BCA}$ entails a further approximation, denoted as Bogoliubov Canonic Approximation (BCA). As shown in Appendix A, BCA follows from assuming $|\:N_{in}\pm2,\:N_{out}\:\rangle\approx|\:N_{in},\:N_{out}\:\rangle$, which is actually the same approximation as SBA, applied in a different context. The common point of weakness is the absence of quantitative control on the order of approximation involved. The validity of SBA and BCA was actually debated through the years, both for demostrating their asymptotic correctness in the TL \cite{G, E}, and for suggesting some corrective strategies, like treating the c-numbers $C$, $C^*$ in eq.n~\eqref{HSBA} as suitable free parameters \cite{HP}. In the recent literature, SBA and BCA seem to be accepted as \emph{bona fide} procedures, without special scrutiny \cite{AB, AB2, PS}. 

In an attempt to check the validity of SBA and BCA, we will study (Section~\ref{DiagHBCA}) the eigenstates $|\:S,\:\mathbf{k}\:\rangle_{BCA}$ and eigenvalues $E_S(k)$ of the Hamiltonian $H_{BCA}$ (eq.n~\eqref{HBCA''}). In Section~\ref{DiagHc}, the problem of the \emph{exact} diagonalization of $H_c$ is approached. A special class of eigenstates $|\:S,\:\mathbf{k}\:\rangle_{c}$ and eigenvalues $\mathcal{E}_S(k)$ are calculated \emph{analytically}. The Fock subspace spanned by such eigensolutions is formed by states with $n_\mathbf{k}=n_{-\mathbf{k}}$, i.e. with the same number of bosons in $|\:\mathbf{k}\:\rangle$ and $|\:-\mathbf{k}\:\rangle$. Due to the symmetry in the exchange $\mathbf{k}\leftrightarrow-\mathbf{k}$, those exact eigenstates correspond to a \emph{vanishing} total momentum, and are denoted as  \textquoteleft s-pseudobosons\textquoteright.  The results of some preliminary calculations are reported, concerning a different class of eigenstates, denoted as \textquoteleft $\eta$-pseudobosons\textquoteright, with asymmetric populations, such that $|n_\mathbf{k}-n_{-\mathbf{k}}|=\eta=1,\:2,\:\dots$ 

In Section~\ref{CompDisc}, it will be shown that $E_S(k)=\mathcal{E}_S(k)$, i.e. the BCA energy eigenvalues coincide with those of the s-eigenstates. Furthermore, $|\:S,\:\mathbf{k}\:\rangle_{c}$ turns out to be eigenstate of the number operator $B_\mathbf{k}^\dagger B_\mathbf{k}$ too, with eigenvalue $S$ (the number of activated pseudobosons). This seems to support the asymptotic correctness of SBA in the TL. In contrast, it will be seen that the pseudobosons created/annihilated by $B_\mathbf{k}^\dagger$ and $B_\mathbf{k}$ do not correspond to the s-eigenstates. Actually, $B_\mathbf{k}^\dagger$ and $B_\mathbf{k}$ project out $|\:S,\:\mathbf{k}\:\rangle_{c}$, into a space \emph{orthogonal} to $|\:S\pm1,\:\mathbf{k}\:\rangle_{c}$. In conclusion: the s-eigenstates $|\:S,\:\mathbf{k}\:\rangle_{c}$ differ substantially from the $|\:S,\:\mathbf{k}\:\rangle_{BCA}$ resulting from BCA, even in the TL. \newline    

An alternative to SBA and BCA is provided by the Gross-Pitaevskii theory (GPT), in which the whole problem of the weakly interacting bosonic gas is turned into a non linear field equation \cite{Gr, Pit}. In Section~\ref{CompDisc}, the expression obtained by GPT, for the \emph{quantum} depletion of the condensate, is shown to coincide exactly with the results deduced from the s-pseudobosons in Section \ref{DiagHc}. In contrast, the \emph{thermal} depletion's formulas differ by terms vanishing with $k=|\mathbf{k}|$, showing the long wavelength nature of the approximations underlying GPT.

Since the s-pseudobosons do not carry a net momentum, Landau's picture of kinetic energy dissipation \cite{L}, for a body flowing in a Bogoliubov superfluid, must involve the emission of $\eta$-pseudobosons, whose detailed features and properties are still under scrutiny. Nevertheless, some preliminary calculations show that the kinematics of the $\eta$-pseudobosons could lead to a new \emph{multichannel} picture of the dissipation processes, significantly different from Landau's theory, based on BCS. This is the program of future investigastions, which are in progress.   

The grand canonic case is briefly discussed in Section~\ref{CompDisc}, in order to compare the present results with the approach developed in ref.s~\cite{AB, AB2}, leading to the Superstable Bogoliubov Hamiltonian.

\section{Low energy eigenstates of $H_{BCA}$}
\label{DiagHBCA}

In ref. \cite{AB} (eq. (2.9)), the canonic Hamiltonian $\widetilde{h}_c$ is expressed in a form reminiscent of \eqref{HSBA}, on applying the approximation $|\:N_{in}\pm2,\:N_{out}\:\rangle\approx|\:N_{in},\:N_{out}\:\rangle$, denoted as BCA (see Appendix A): 

\begin{subequations}
\begin{align}
H_{BCA}&=E_{in}+\sum_{\mathbf{k}\ne0}\overbrace{\left[\mathcal{T}(k)+N\:\widehat{u}(k)\right]}^{\epsilon_1(k)}\beta^\dagger_{\mathbf{k}}\beta_{\mathbf{k}}+\nonumber\\
&+\frac{N}{2}\sum_{\mathbf{k}\ne0}\widehat{u}(k)\Big[\beta^\dagger_{\mathbf{k}}\beta^\dagger_{-\mathbf{k}}+\beta_{\mathbf{k}}\beta_{-\mathbf{k}}\Big]\:,\label{HBCA}
\end{align}
\\
where new creation/annihilation operators are introduced:

\begin{equation}
\label{beta,beta*}
\beta_{\mathbf{k}}=b_{\mathbf{0}}^\dagger\left(\widetilde{N}_{in}+1\right)^{-1/2}b_\mathbf{k}\quad,\quad\beta_{\mathbf{k}}^\dagger=b_{\mathbf{k}}^\dagger\left(\widetilde{N}_{in}+1\right)^{-1/2}b_\mathbf{0}\:,
\end{equation}
\end{subequations}
\\
which ensure the conservation of the number $N$ of real bosons. Note that $\beta_\mathbf{k}$ and $\beta_\mathbf{k}^\dagger$ exactly satisfy the canonic commutation rules (CCR)\footnote{Notice that $F(\widetilde{N}_{in})\beta_{\mathbf{k}}=\:\beta_{\mathbf{k}}F(\widetilde{N}_{in}-1)$ for any function $F$.}. The next step are the well known Bogoliubov transformations:

\begin{subequations}
\begin{equation}
\label{defB}
B_{\mathbf{k}}=w_+(\mathbf{k})\:\beta_\mathbf{k}-w_-(\mathbf{k})\:\beta^\dagger_{-\mathbf{k}}\quad;\quad B^\dagger_{\mathbf{k}}=w_+^{\:*}(\mathbf{k})\beta^\dagger_{\mathbf{k}}-w_-^{\:*}(\mathbf{k})\beta_{-\mathbf{k}}\:,
\end{equation}
\\
according to which an appropriate choice of $w_\pm(\mathbf{k})$ leads to the non interacting form eq.n~\eqref{HBCA''}. The one-momentum Hamiltonian $\widetilde{h}_{BCA}(\mathbf{k})$ has eigenvalues (recall eq.n~\eqref{HFBA} and the definition of $\epsilon_1(k)$ in eq.n~\eqref{HBCA}):

\begin{equation}
\label{ES1}
E_{S}(k)=\epsilon(k)\left(S+\frac{1}{2}\right)-\frac{\epsilon_1(k)}{2}\quad(S=0,\:1,\:\dots)\:,
\end{equation}
\\
with \cite{PS}:

\begin{align}
\label{epsilonBog}
\epsilon(k)&=\sqrt{\mathcal{T}^2(k)+2N\:\widehat{u}(k)\mathcal{T}(k)}=\nonumber\\
\nonumber\\
&=\frac{\hbar\:k}{\sqrt{2M}}\sqrt{2N\:\widehat{u}(k)+\frac{\hbar^2k^2}{2M}}\:.
\end{align}
\\
and:

\begin{equation}
\label{wpm}
w_\pm(\mathbf{k})=\pm\sqrt{\frac{\epsilon_1(k)}{2\epsilon(k)}\pm\frac{1}{2}}\:.
\end{equation}
\end{subequations}
\\
Apart from the rigorous definitions \eqref{beta,beta*} \cite{AB}, what precedes is a standard issue, currently reported, with minor changes, in several works and textbooks\footnote{In most cases, the creation/annihilation operators are defined as $\beta_\mathbf{k}^\dagger=b_\mathbf{0}b_\mathbf{k}^\dagger/\sqrt{N}$, and $\beta_\mathbf{k}=b_\mathbf{0}^\dagger b_\mathbf{k}/\sqrt{N}$, which satisfies the CCR only if $n_\mathbf{k}$ is a sub-extensive quantity.}. What is lacking, to the author's knowledge, is a concrete representation of the pseudobosons created/annihilated by  $B^\dagger_{\mathbf{k}}$, $B_{\mathbf{k}}$. A way to approach the problem is finding the eigenstates of $H_{BAC}$, corresponding to $S$ pseudobosons, as:

\begin{equation}
\label{| S >BCA}
|\:S,\:\mathbf{k}\:\rangle_{BCA}=\frac{(B_\mathbf{k}^\dagger)^S}{\sqrt{S!}}|0,\:\mathbf{k}\:\rangle_{BCA}\:,
\end{equation}
\\
in terms of the $k$-pseudobosons \textquoteleft vacuum\textquoteright$\:|0,\:\mathbf{k}\:\rangle_{BCA}$, defined by the basic condition:

\begin{equation}
\label{VacBCA}
B_\mathbf{k}|0,\:\mathbf{k}\:\rangle_{BCA}=0\:.
\end{equation}
\\
Let us deal with the subspace spanned by the $N$-particle Fock states with $j$ (real) bosons occupying $|\:\mathbf{-k}\:\rangle$,  $(j+\eta)$ bosons occupying $|\:\mathbf{k}\:\rangle$ and $(N-2j-\eta)$ occupying $|\:\mathbf{0}\:\rangle$:

\begin{subequations}
\begin{equation}
\label{| j >eta}
|\:j,\:\mathbf{k}\:\rangle_\eta=\frac{(b_\mathbf{0}^\dagger)^{N-2j-\eta}}{\sqrt{(N-2j-\eta)!}}\frac{(b_\mathbf{k}^\dagger)^{j+\eta}(b_\mathbf{-k}^\dagger)^j}{\sqrt{j!(j+\eta)!}}|\emptyset\:\rangle\:,
\end{equation}
\\
where $\eta=0,\:1,\:\dots$\footnote{The same procedure can be applied to states with $j$ bosons occupying $|\:\mathbf{k}\:\rangle$,  $(j+\eta)$ bosons occupying $|\:\mathbf{-k}\:\rangle$, which would be equivalent to change the sign of $\eta$.}. We guess a possible form of the vacuum $|0,\:\mathbf{k}\:\rangle_{BCA}$ as follows:

\begin{equation}
\label{| 0 >BCA}
|\:0,\:\mathbf{k}\:\rangle_{BCA}=\sum_{j=0}^{M}\phi_0(j)|\:j,\:\mathbf{k}\:\rangle_\eta\:,
\end{equation}
\end{subequations}
\\
with $M\ll N/2$ and $N$ the conserved number of bosons. On account of eq.ns~\eqref{beta,beta*}, \eqref{defB}, the condition~\eqref{VacBCA} leads to the following equation:

\begin{align}
&\phi_0(0)\left[w_+\sqrt{\eta}|\:0,\:\mathbf{k}\:\rangle_{\eta-1}-w_-|\:1,\:\mathbf{k}\:\rangle_{\eta-1}\right]+\label{0}\\
+&\phi_0(1)\left[w_+\sqrt{\eta+1}|\:1,\:\mathbf{k}\:\rangle_{\eta-1}-w_-\sqrt{2}|\:2,\:\mathbf{k}\:\rangle_{\eta-1}\right]+\dots\nonumber\\
+&\phi_0(j)\left[w_+\sqrt{\eta+j}|\:j,\:\mathbf{k}\:\rangle_{\eta-1}-w_-\sqrt{j+1}|\:j+1,\:\mathbf{k}\:\rangle_{\eta-1}\right]+\nonumber\\
+&\phi_0(j+1)\left[w_+\sqrt{\eta+j+1}|\:j,\:\mathbf{k}\:\rangle_{\eta-1}-w_-\sqrt{j+2}|\:j+2,\:\mathbf{k}\:\rangle_{\eta-1}\right]+\nonumber\\
+&\dots=0\:,\label{all}
\end{align}
\\
that can be solved by equating the second term in each line with the first one in the next line. However, this forces the first term of the first line to vanish, i.e. $w_+\sqrt{\eta}|\:0,\:\mathbf{k}\:\rangle_{\eta-1}=0$, whose solution implies $\eta=0$. The resulting recurrence formula is trivially solved by $\phi_0(j)=\left(w_-/w_+\right)^j\phi_0(0)$. From eq.n~\eqref{wpm}, it follows that $|w_-/w_+|<1$, hence, for $M\gg1/\ln(|w_+/w_-|)$, one has (eq.n~\eqref{| 0 >BCA}):

\begin{subequations}
\begin{equation}
\label{| 0 >BCA'}
|\:0,\:\mathbf{k}\:\rangle_{BCA}=\phi_0(0)\sum_{j=0}^{\infty}\left[\frac{w_-(k)}{w_+(k)}\right]^j|\:j,\:\mathbf{k}\:\rangle=|\:0,\:-\mathbf{k}\:\rangle_{BCA}\:,
\end{equation}
\\
with:

\begin{equation}
\label{| j >}
|\:j,\:\mathbf{k}\:\rangle\equiv|\:j,\:\mathbf{k}\:\rangle_{\eta=0}=\frac{(b_\mathbf{0}^\dagger)^{N-2j}}{\sqrt{(N-2j)!}}\frac{(b_\mathbf{k}^\dagger)^j(b_\mathbf{-k}^\dagger)^j}{j!}|\emptyset\:\rangle\:.
\end{equation}
\end{subequations}
\\
From eq.n~\eqref{| 0 >BCA}, a straightforward calculation yields the normalized 1-pseudoboson states, corresponding to $\mathbf{k}$ and $-\mathbf{k}$:

\begin{subequations}
\label{|1,k>BCA}
\begin{align}
|\:1,\:\pm\mathbf{k}\:\rangle_{BCA}&=B^\dagger_{\pm\mathbf{k}}|0,\:\mathbf{k}\:\rangle_{BCA}=\nonumber\\
\nonumber\\
&=\frac{1}{w_-(k)}\sum_{j=1}^{\infty}\sqrt{j}\left[\frac{w_-(k)}{w_+(k)}\right]^j|\:j-1,\:\pm\mathbf{k}\:\rangle_{1}\:,
\end{align}
\\
where, according to eq.n~\eqref{| j >eta}, one has:

\begin{equation}
|\:j-1,\:-\mathbf{k}\:\rangle_{1}=|\:j,\:\mathbf{k}\:\rangle_{-1}\:.
\end{equation}
\end{subequations}
\\
Given the total momentum operator $\mathbf{P}_{\mathbf{k}}=\hbar\mathbf{k}[\widetilde{n}_\mathbf{k}-\widetilde{n}_{-\mathbf{k}}]$, it is easily seen that:

\begin{equation}
\label{<Pk>}
\mathbf{P}_\mathbf{k}|\:S,\:\pm\mathbf{k}\:\rangle_{BCA}=\pm S\hbar\mathbf{k}|\:S,\:\pm\mathbf{k}\:\rangle_{BCA}\:,
\end{equation}
\\
showing that the state of  $S$ BCA pseudobosons carries a momentum $S\hbar\mathbf{k}$.

\section{Exact low energy eigenstates of $H_{c}$: s- and $\eta$-pseudobosons}
\label{DiagHc}

Since $\mathcal{T}(k)$ and $\widehat{u}(k)$ depend on $k=|\:\mathbf{k}\:|$, the Hamiltonian $\widetilde{h}_c$ (eq.n~\eqref{Hc}) can be expressed as a sum of independent one-momentum Hamiltonians

\begin{subequations}
\begin{equation}
\label{Hc2}
H_{c}=E_{in}+\sum_{\mathbf{k}\ne0}\widetilde{h}_c(\mathbf{k})\:,
\end{equation}
\\
where:

\begin{align}
\widetilde{h}_c(\mathbf{k})&=\frac{1}{2}\widetilde{\epsilon}_1(k)[b^\dagger_{\mathbf{k}}b_{\mathbf{k}}+b^\dagger_{-\mathbf{k}}b_{-\mathbf{k}}]+\nonumber\\
\nonumber\\
&+\frac{1}{2}\widehat{u}(k)\Big[b^\dagger_{\mathbf{k}}b^\dagger_{-\mathbf{k}}(\:b_{\mathbf{0}}\:)^2+b_{\mathbf{k}}b_{-\mathbf{k}}(\:b^\dagger_{\mathbf{0}}\:)^2\Big]\:.\label{htilde}
\end{align}
\end{subequations}
\\

In the present section we study the \emph{exact} eigenstates of $H_{c}$ (eq.n~\eqref{htilde}), starting with the Fock subspace spanned by the states eq.n~\eqref{| j >}, with $j$ (real) bosons occupying $|\:\pm\mathbf{k}\:\rangle$ and $N-2j$ occupying $|\:\mathbf{0}\:\rangle$. To remind the symmetric nature of the $|S,\:\mathbf{k}\:\rangle_c$'s, for $\mathbf{k}\leftrightarrow-\mathbf{k}$, we call them \textquoteleft s-eigenstates\textquoteright$\:$ (or \textquoteleft s-pseudobosons\textquoteright). We guess a possible form of the $N$-particle s-eigenstate corresponding to a given momentum $\hbar\mathbf{k}$ as follows:

\begin{equation}
\label{| S >}
|\:S,\:\mathbf{k}\:\rangle_{c}=\sum_{j=0}^{M}\phi_S(j)|\:j,\:\mathbf{k}\:\rangle\:,
\end{equation}
\\
with $M\ll N/2$ and $N$ the conserved number of bosons. The coefficients $\phi_S(j)$ are, obviously, the unknowns of the problem. The index $S=0,\:1,\:\dots$ labels the energy eigenvalues, as we shall see in what follows. From eqn.~\eqref{Hc2}, the eigenvalue equation becomes:

\begin{equation}
\label{EigenEq}
\widetilde{h}_c(\mathbf{k})|\:S,\:\mathbf{k}\:\rangle_{c}=\mathcal{E}_S(\mathbf{k},\:N)|\:S,\:\mathbf{k}\:\rangle_{c}\:.
\end{equation}
\\
In the following calculations, we drop the dependence on $\mathbf{k}$ and $N$ if not necessary, and set:

\begin{equation*}
\frac{m}{N}=\delta_m\:\:(m=0,\:1,\:\dots).
\end{equation*}
\\
With that convention, equations~\eqref{Hc2} - \eqref{EigenEq} yield:

\begin{align*}
&\widetilde{h}_c(\mathbf{k})|\:j,\:\mathbf{k}\:\rangle=j\:\left[\mathcal{T}+N\widehat{u}(1-\delta_{2j})\right] |\:j,\:\mathbf{k}\:\rangle+\nonumber\\
\nonumber\\
&+\frac{N\widehat{u}}{2}\Big[|\:j+1,\:\mathbf{k}\:\rangle(j+1)\sqrt{(1-\delta_{2j})(1-\delta_{2j+1})}+\nonumber\\
\nonumber\\
&+|\:j-1,\:\mathbf{k}\:\rangle\: j\:\sqrt{(1-\delta_{2j-1})(1-\delta_{2j-2})}\:\Big]\:.
\end{align*}
\\
Let the upper value $M$ in the sum \eqref{| j >} be a \emph{subextensive diverging} quantity, i.e. $\lim_{V\rightarrow\infty}M/V=0$, $\lim_{V\rightarrow\infty}M=\infty$. In the TL, this makes it possible to have an arbitrarly large $M$ in the sum, and all the $\delta_m$'s vanishing in the preceding equation\footnote{Since it will be shown that the leading term of $|\phi_S(j)|$ (eq.n~\eqref{| j >}) is proportional to $j^S\mathrm{e}^{-j\gamma}$, one could actually take for $M$ a finite value, large compared to $S/\gamma$.}, which yields:

\begin{align*}
&\widetilde{h}_c(\mathbf{k})|\:j,\:\mathbf{k}\:\rangle=\epsilon_1\:j|\:j,\:\mathbf{k}\:\rangle+\nonumber\\
\nonumber\\
&+\frac{N\widehat{u}}{2}\Big[|\:j+1,\:\mathbf{k}\:\rangle(j+1)+|\:j-1,\:\mathbf{k}\:\rangle\:j\:\Big]\:.
\end{align*}
\\
By means of the preceding equation, the eigenvalue equation \eqref{EigenEq} reads:

\begin{subequations}
\begin{align}
&\left[\underline{\epsilon}_1\:m-\underline{\mathcal{E}}_S\right]\phi_S(m)+\nonumber\\
\nonumber\\
&+\frac{1}{2}[\phi_S(m+1)(m+1)+m\:\phi_S(m-1)]=0\:,\label{EigenEq2}
\end{align}
\\
where:

\begin{equation}
\label{A}
\underline{A}=\frac{A}{N\widehat{u}}\:,
\end{equation}
\end{subequations}
\\
for each of the quantities $A=\mathcal{E}_S,\:\epsilon_1,\:\epsilon$. Equation \eqref{EigenEq2} can be easily transformed as:

\begin{subequations}
\label{EigenEq3All} 
 \begin{equation}
 \phi_S(m)[D+Bm]+\phi_S(m-1)m+\phi_S(m+1)(m+1)=0 \:,\label{EigenEq3}
 \end{equation}
\\
on setting 

\begin{equation}
\label{m,D,B}
D=-2\underline{\mathcal{E}}_S\quad;\quad B=2\underline{\epsilon}_1\:.
\end{equation}
\end{subequations}
\\
Now, let:

\begin{equation}
\phi_S(m)=x^m\underbrace{\sum_{s=0}^SC_s\:m^s}_{P_S(m)}\:, \label{phiS}
\end{equation}
\\
where $x$ and the $C_s$'s are the unknowns to be determined. The boundary conditions are the normalizability of $|\:S,\:\mathbf{k}\:\rangle_{c}$, and the absence of negative occupation numbers, that yields, in this case, $\phi_S(-1)=0$. On account of eq.n \eqref{phiS}, equation~\eqref{EigenEq3} becomes:

\begin{equation*}
[D+Bm]P_S(m)+\frac{m}{x}P_S(m-1)+x(m+1)P_S(m+1)=0\:.
\end{equation*}
\\
The l.h.s. of the preceding equation is a (S+1)-degree polinomial in $m$. The solution then follows from a system of $S+2$ equations, each corresponding to the vanishing of the coefficient of $m^l$, with $l=0,\:1,\:\dots,\:S+1$\footnote{A further equation follows from the normalization of $|\:S,\:\mathbf{k}\:\rangle_{c}$. Note that the whole eigenvalue problem has $S+3$ unknowns (the $C_s$'s, $x$ and $D$, which contains the eigenvalue $\mathcal{E}_S$).}:

\begin{align}
&DC_l+ BC_{l-1}+x\left[\sum_{s=l}^SC_s\binom{s}{l}+\sum_{s=l-1}^SC_s\binom{s}{l-1}\right]-\nonumber\\
\nonumber\\
&-\frac{1}{x}\sum_{s=l-1}^SC_s\binom{s}{l-1}(-1)^{s-l}=0\quad(l=0,\:1,\:\dots\:,S+1)\label{EqP}
\end{align}
\\
(with $C_{S+1}=0$ by definition). The vanishing of the two terms $l=S+1$ and $l=S$ is sufficient to determine the two unknowns $D$ and $x$, i.e. the eigenvalue and the exponential slope $\pm\ln(|x|)$ of the eigenstate (recall eq.ns~\eqref{m,D,B} and \eqref{phiS}). Actually, equation~\eqref{EqP} yields:

\begin{subequations}
\label{x,ES}
\begin{equation}
\label{x}
 B+x+x^{-1}=0\quad(l=S+1) \:, 
\end{equation}
\\ 
 whence: 
 
 \begin{equation}
 \label{ES}
 D+x +S(x-x^{-1})=0\quad(l=S)\:.
 \end{equation}
 \\
Note that both $x$ and $D$ are independent from the $C_s$'s, which are determined by the next equations~\eqref{EqP} and by normalization. In particular, it is important to explicitate the case $l=0$, which implies:

\begin{equation}
\label{EqP0}
DC_0+x\sum_{s=0}^{S}C_{s}=0\:.
\end{equation}
 \end{subequations}
 \\
For the state $|\:S,\:\mathbf{k}\:\rangle_{c}$ (eq.n~\eqref{| j >}) to be normalizable, the solution $x$ of the 2nd degree equation \eqref{x} must be smaller than 1 in modulus. Recalling eq.ns~\eqref{m,D,B} and \eqref{epsilonBog}, one finally gets:

\begin{subequations}
\label{x,1/x}
\begin{align}
x&=\overbrace{\sqrt{\underline{\epsilon}_1^2(k)-1}}^{\underline{\epsilon}(k)}-\underline{\epsilon}_1(k)=\underline{\epsilon}(k)-\sqrt{\underline{\epsilon}^2(k)+1}\label{x(k)}\\
\nonumber\\
\frac{1}{x}&=-\left[\sqrt{\underline{\epsilon}_1^2(k)-1}+\underline{\epsilon}_1(k)\right]=-\left[\underline{\epsilon}(k)+\sqrt{\underline{\epsilon}^2(k)+1}\right]\:,\label{1/x(k)}
\end{align}
\\
since $x^{-1}$ is the other solution. With the aid of eq.ns~\eqref{m,D,B}, \eqref{A}, \eqref{epsilonBog}, equation~\eqref{ES} yields:

\begin{align}
\mathcal{E}_S(k)&=\frac{\widehat{u}(k)N}{2}\:x+S\:\epsilon(k)=\epsilon(k)\left(S+\frac{1}{2}\right)-\frac{\epsilon_1(k)}{2}=\nonumber\\
\nonumber\\
&\frac{\hbar\:k}{\sqrt{2M}}\sqrt{2N\:\widehat{u}(k)+\frac{\hbar^2k^2}{2M}}\left(S+\frac{1}{2}\right)-\frac{\epsilon_1(k)}{2}=E_S(k)\:,\label{ES(k)}
\end{align}
\end{subequations}
\\
with $k$ and $N$ restored everywhere. Since all the unknowns of the problem are determined at the present stage, one might wonder what about the boundary condition $\phi_S(-1)=0$, that means

\begin{equation}
\label{BC}
DP_S(0)+xP_S(1)=0\:,
\end{equation}
\\
according to eq.n~\eqref{phiS}. However, equation~\eqref{BC} turns out to be the same as \eqref{EqP0}. Therefore, solving the system of equations \eqref{EqP} means satisfying the boundary condition eq.n~\eqref{BC} too.

A case of special interest is the vacuum $|\:0,\:\mathbf{k}\:\rangle_c$ of the s-pseudobosons, that follows from eq.ns~\eqref{| S >}, \eqref{phiS} with $S=0$:

\begin{subequations}
\label{| 0 >c}

\begin{equation}
|\:0,\:\mathbf{k}\:\rangle_c=C_0\sum_{j=0}^\infty\:x^j\:|\:j,\:\mathbf{k}\:\rangle=C_0\sum_{j=0}^\infty\:\left[\frac{w_-(k)}{w_+(k)}\right]^j\:|\:j,\:\mathbf{k}\:\rangle\:,
\end{equation}
\\
where the second equality follows from eq.ns~\eqref{wpm}, \eqref{x,1/x}, and:

\begin{equation}
C_0=\sqrt{1-x^2}=\sqrt{1-|w_-/w_+|^2}\:.
\end{equation} 
\end{subequations}
\\
In analogy with what has been done in Section~\ref{DiagHBCA}, it is useful to express the exact single s-pseudoboson state:

\begin{subequations}
\label{s-states}
\begin{equation}
\label{|1,k>c}
|1,\:\mathbf{k}\:\rangle_c=\frac{1}{1-x^2}\sum_{m=0}^\infty x^m[x^2-m(1-x^2)]\:|\:m,\mathbf{k}\:\rangle\:.
\end{equation}
\\

Due to the symmetry in the populations of excited bosons with opposite momenta, a straightforward consequence of eq.n~\eqref{| S >} is the vanishing of the total momentum carried by the s-eigenstates:

\begin{equation}
\label{Pk=0}
\mathbf{P}_\mathbf{k}|S,\:\mathbf{k}\:\rangle_c=0\:.
\end{equation}
\end{subequations}
\\

It should be clear that the s-eigenstates, described so far, do not form a base: they just span a Fock subspace, orthogonal to all states like $|j,\:\mathbf{k}\:\rangle_\eta$, defined by eq.n~\eqref{| j >eta}, with $j$ bosons in $|\:-\mathbf{k}\:\rangle$ and $j+\eta$ bosons in $|\:\mathbf{k}\:\rangle$. The diagonalization of $\widetilde{h}_c$ in the non symmetric subspace is far from trivial. For each $\eta>0$, one could guess the form of the eigenstate as:

\begin{subequations}
\begin{equation}
\label{| S >eta}
|S,\:\mathbf{k},\:\eta\:\rangle_c=\sum_{j=0}^\infty\phi_S(j,\eta)|j,\:\mathbf{k}\:\rangle_\eta\:,
\end{equation}
\\
in analogy with eq.n~\eqref{| S >}, and solve the eigenvalue equation

\begin{equation}
\label{DiagHceta}
\widetilde{h}_c|S,\:\mathbf{k},\:\eta\:\rangle_c=\mathcal{E}_S(k,\eta)|S,\:\mathbf{k},\:\eta\:\rangle_c
\end{equation}
\end{subequations}
\\
in the unknowns $\phi_S(j,\eta)$. In what follows, the eigenstates eq.n~\eqref{| S >eta}, with a population asymmetry $\eta$, will be defined \textquoteleft $\eta$-eigenstates\textquoteright (or \textquoteleft $\eta$-pseudobosons\textquoteright). A preliminary result of calculatios that are in progress, is that, for \emph{finite} values of the population asymmetry $\eta$, the condition~\eqref{x} remains the same, i.e., the exponential factor $x^j$, ensuring the normalizability, does not change. At present, however, there is no demonstration that the label $S$ does numerate the pseudobosons, as shown for the s-eigenstates. Actually, it is immediately seen that:

\begin{equation}
\label{Pketa}
\mathbf{P}_\mathbf{k}|S,\:\mathbf{k},\:\eta\:\rangle_c=\eta\hbar\mathbf{k}|S,\:\mathbf{k},\:\eta\:\rangle_c\:,
\end{equation}
\\
i.e. the total momentum of the $\eta$-eigenstates corresponds to $\eta$ free particles. Hence, one might suspect that the number of pseudobosons contained in $|S,\:\mathbf{k},\:\eta\:\rangle_c$ is $\eta$ and not $S$. This non trivial problem is under scrutiny, and the results will hopefully appear in a forthcoming paper. However, the dependence of the energy on \emph{two} indices $S$ and $\eta$ has, by itself, important consequences, that will be discussed in the next section.

\section{Comparisons and discussion}
\label{CompDisc}

The results expressed by eq.ns~\eqref{ES(k)} and \eqref{| 0 >c} are noteworthy: the \emph{exact} eigenvalues $\mathcal{E}_S(k)$, corresponding to the s-eigenstates $|\:S,\:\mathbf{k}\:\rangle_{c}$, are identical to the energies $E_S(k)$, obtained in Section~\ref{DiagHBCA} (eq.ns~\eqref{HBCA''}), from the BCA. The vacuum of the s-pseudobosons (eq.n~\eqref{| 0 >c}) is the same as the one calculated from BCA (eq.n~\eqref{| 0 >BCA}). Furthermore, from a straightforward calculation it follows that: 

\begin{equation}
\label{NpseudoOK}
B_\mathbf{k}^\dagger B_\mathbf{k}|\:S,\:\mathbf{k}\:\rangle_{c}=S|\:S,\:\mathbf{k}\:\rangle_{c}
\end{equation}
\\
in the TL, which shows that the s-eigenstates of the Hamiltonian $\widetilde{h}_c(\mathbf{k})$ are eigenstates of the BCA number operator too. In spite of this tight correspondence, however, what follows from Section~\ref{DiagHc} displays differences from BCA, that do not vanish in the TL and cannot be neglected. The s-pseudobosons are created/annihilated by enhancing/diminishing the number of terms in the polinomial $P_S(m)$ (eq.n~\eqref{phiS}), so that their number coincides with the degree $S$ of the polinomial itself. From eq.n~\eqref{NpseudoOK}, one might expect that this procedure is equivalent to apply $B_\mathbf{k}^\dagger$ and $B_\mathbf{k}$ to the s-eigenstates, i.e.:

\begin{equation*}
\left.
\begin{array}{l}
|B_\mathbf{k}^\dagger |\:S,\:\mathbf{k}\:\rangle_{c}=\sqrt{S+1}|\:S+1,\:\mathbf{k}\:\rangle_{c}\\
\\
B_\mathbf{k}|\:S,\:\mathbf{k}\:\rangle_{c}=\sqrt{S}|\:S-1,\:\mathbf{k}\:\rangle_{c}
\end{array}\right\}\text{ (wrong)}\:,
\end{equation*}
\\
but this is definitely not the case, instead. Actually, it is easily seen that:

\begin{equation}
\label{ortho}
_c\langle\:\mathbf{k}\:, S+1\:| B_\mathbf{k}^\dagger|\:S,\:\mathbf{k}\:\rangle_{c}=\:_c\langle\:\mathbf{k}\:, S-1\:| B_\mathbf{k}|\:S,\:\mathbf{k}\:\rangle_{c}=0\:,
\end{equation}
\\
which shows that $B_\mathbf{k}^\dagger$ and $B_\mathbf{k}$ project out the s-pseudobosons states, into a space \emph{orthogonal} to $|\:S\pm1,\:\mathbf{k}\:\rangle_{c}$. This is because $B_\mathbf{k}^\dagger$ and $B_\mathbf{k}$ are linear combinations of $b_{\pm\mathbf{k}}^\dagger$ and $b_{\pm\mathbf{k}}$, so that the application to a state $|\:j,\mathbf{k}\:\rangle$ with the \emph{same} number of particles in $|\:\pm\mathbf{k}\:\rangle$ (eq.n~\eqref{| j >}) results in a linear combination of states $|\:j,\mathbf{k}\:\rangle_{\pm1}$ (eq.n~\eqref{| j >eta}), with a \emph{different} number of particles. Obviously, one has $\langle\:\mathbf{k},\:j'\:|\:j,\:\mathbf{k}\:\rangle_{\pm1}=0$, which implies eq.n~\eqref{ortho}.

In short: the BCA quantities, resulting from the number operator $B_\mathbf{k}^\dagger B_\mathbf{k}$ only, are \emph{exact}. However, the \emph{separate} effects of $B_\mathbf{k}^\dagger$ and $B_\mathbf{k}$ are quite different from the \textquoteleft creation/annihilation\textquoteright$\:$of the s-pseudobosons: the eigenstates $|\:S,\:\mathbf{k}\:\rangle_{c}$ and $|\:S,\:\mathbf{k}\:\rangle_{BCA}$ (eq.n~\eqref{| S >BCA}) are quite different, even in the TL.\\

A further fruitful comparison can be done, with the Gross-Pitaevskii theory (GPT)~\cite{Gr, Pit}, that provides an approach alternative to FBA and SBA. In ref.~\cite{PS}, the concentration of \emph{real} excited bosons ($\mathbf{k}\ne\mathbf{0}$), at zero and finite temperature $T$, is calculated, and referred to as \textquoteleft quantum depletion\textquoteright$\:$and \textquoteleft thermal depletion\textquoteright$\:$ of the condensate, respectively. According to the s-pseudobosons formalism, the depletion of the condensate reads:

\begin{equation}
\langle\:N_{out}\:\rangle_T=\underbrace{\sum_{\mathbf{k}\ne\mathbf{0}}\langle\:n_\mathbf{k}\:\rangle_0}_{\text{quantum depletion}}+\overbrace{\sum_{\mathbf{k}\ne\mathbf{0}}\langle\:n_\mathbf{k}\:\rangle_T}^{\text{thermal depletion}}\:,
\end{equation}
\\
where $\langle\:n_\mathbf{k}\:\rangle_0$ and $\langle\:n_\mathbf{k}\:\rangle_T$ are the numbers of \emph{real} excited bosons contained in each state $|\:0,\:\mathbf{k}\:\rangle_{c}$ and $|\:S(T,\:k),\:\mathbf{k}\:\rangle_{c}$, respectively. According to eq.ns \eqref{| S >}, \eqref{phiS} and \eqref{wpm}, one gets:
 
\begin{align}
\label{< n(k) >0}
\langle\:n_\mathbf{k}\:\rangle_0&=\frac{\sum_{j=0}^\infty x^{2j}(k)\:j}{\sum_{j=0}^\infty x^{2j}(k)}=\frac{x^2(k)}{1-x^2(k)}=w_-^2(k)\:.
\end{align}
\\
It can be seen that eq.n~\eqref{< n(k) >0} is exactly the same as the quantum depletion term in eq.n (4.58) of ref.~\cite{PS}, since $w_-(k)=v_{-\mathbf{p}}$ (and $w_+(k)=u_{\mathbf{p}}$; recall eq.ns~\eqref{defB}). On setting:

\begin{equation}
\label{S(T,k)}
S(T,\:k)=\frac{1}{\mathrm{e}^{\beta\epsilon(k)}-1}
\end{equation}
\\
for the thermal value of the number of s-pseudobosons, the thermal depletion can be calculated accordingly, though in this case the comparison with GPT is less straightforward. A lenghty calculation (Appendix B) makes it possible to express $\langle\:n_{\mathbf{k}}\:\rangle_T$ in terms of the Hurwitz-Lerch Phi functions: 

\begin{subequations}
\label{H-L-Phi}
\begin{align}
&\Phi(\lambda,\:-m,\:a)=\sum_{j=0}^\infty \lambda^{2j}(j+a)^{m}=\label{Phi}\\
\nonumber\\
&=\frac{1}{1-\lambda}\sum_{j=0}^m\binom{m}{j}\:a^{m-j}\sum_{r=0}^j\left\{
\begin{array}{c}
j\\
r
\end{array}\right\}\:r!\left(\frac{\lambda}{1-\lambda}\right)^r=\label{serie}\\
\nonumber\\
&=\frac{m!\:\lambda^m}{(1-\lambda)^{m+1}}\left[1+\mathrm{o}(1-\lambda)\right]\:,
\end{align}
\end{subequations}
\\
where $|\lambda|\ <1$, $|a|\ne0$ and $\left\{
\begin{array}{c}
j\\
r
\end{array}\right\}$ are Stirling's numbers of the second rank (see eq.n (6.3) in ref.~\cite{B}). The result of interest for the present comparison yields $\langle\:n_{\mathbf{k}}\:\rangle_T$ to the leading order in $|1-x^2|^{-1}$ and $S^{-1}$ (Appendix B):

\begin{subequations}
\begin{align}
\langle\:n_{\mathbf{k}}\:\rangle_T&=\sum_{j=0}^\infty x^{2j}\:j\left[\sum_{s=0}^SC_s\:j^s\right]^2=\label{< n(k) >T}\\
\nonumber\\
&=\frac{2\:S(T,\:k)}{1-x^2(k)}\left[1+\mathrm{o}(1-x^2)+\mathrm{o}(S^{-1})\right]\:.\label{< n(k) >T'}
\end{align}
\end{subequations}
\\
From eq.ns~\eqref{wpm}, and \eqref{x,1/x}, one gets:

\begin{equation}
w_+^2(k)+w_-^2(k)=\frac{1+x^2(k)}{1-x^2(k)}=\frac{2}{1-x^2(k)}\left[1+\mathrm{o}(1-x^2)\right]\:,
\end{equation}
\\
to the same leading order. Since  $w_+^2(k)+w_-^2(k)=u^2_\mathbf{p}+w^2_{-\mathbf{p}}$, from eq.n~\eqref{S(T,k)} it is seen that the thermal depletion in eq.n (4.58) of ref.~\cite{PS} corresponds to eq.n~\eqref{< n(k) >T}, \emph{modulo} terms small to order $1-x^2$ and $S^{-1}$. This means that GPT follows from a long wavelength approximation. In fact, recalling eq.ns~\eqref{x(k)}, \eqref{epsilonBog}, \eqref{A} and eq.n~\eqref{S(T,k)}, it easily seen that:

\begin{align}
1-x^2(k)=2\underline{\epsilon}(k)\left[1+\mathrm{o}(\underline{\epsilon})\right]&=\frac{2\:\hbar\:k}{\sqrt{N\widehat{u}(\mathbf{0})\:M}}\left[1+\mathrm{o}(k)\right]\nonumber\\
\nonumber\\
S^{-1}(T,\:k)=\beta\epsilon(k)\left[1+\mathrm{o}(\beta\epsilon)\right]&= \frac{\:\hbar\:k}{\kappa T}\sqrt{\frac{N\widehat{u}(\mathbf{0})}{M}}\left[1+\mathrm{o}(k)\right]\nonumber\:,
\end{align}
\\
which shows that neglecting $1-x^2(k)$ and $1/S(T,\:k)$ is a small-$k$ approximation. In summary, the \emph{quantum} depletion of the condensate, deduced from GPT, is exact, while the \emph{thermal} depletion applies to long wavelength pseudobosons only. \newline

The differences described above, between BCA and exact pseudobosons, have further consequences too, that could reflect in the detailed dynamics of the dissipation processes. In general, BCA results in a \textquoteleft standard\textquoteright$\:$ picture of massless bosons, since eq.ns~\eqref{<Pk>} show that each BCA pseudoboson carries a total momentum $\hbar\mathbf{k}$ and an energy $\epsilon(k)$. This picture underlies Landau's theory of dissipation, in a Bogoliubov gas, as due to the emission of a single pseudophonon, satisfying the energy/momentum conservation, from a body flowing in the superfluid \cite{L}. The process just outlined is impossible for the s-pseudobosons, since their total momentum is zero (eq.n~\eqref{Pk=0}), though their energy is $(S+1/2)\epsilon(k)$, with arbitrary $S=1,\:,2,\:\dots$. So, there exists a class of \emph{exact} pseudobosons that can influence the thermodynamics of the gas (in particular, the condensate depletion), but do not enter the dissipation processes. The only possibility for Landau's picture of dissipation to apply is thereby the emission of $\eta$-pseudobosons. However, the dependence of their energy on $S$ and $\eta$ (eq.n~\eqref{DiagHceta}), leads one to suspect that, unlike BCA pseudobosons, the dissipation processes involving the $\eta$-pseudobosons could have many emission channels, corresponding to any possible change with $S$ of the energy $\mathcal{E}_S(k,\:\eta)$, at fixed momentum $\eta\hbar\mathbf{k}$. If more accurate calculations (that are in progress) will confirm what precedes, there would be important consequences for the dissipation dynamics.\newline

The \emph{grand canonic} case shows some controvesial aspects not discussed in the present work. In ref.s \cite{AB, AB2} the instability of the Hamiltonian $H_{FBA}$ (eq.n~\eqref{HFBA}) for positive chemical potentials, and the possible existence of a gap in the energy spectrum, have been stressed as the main points of weakness of FBA. The suggestion is changing the truncation of the interaction terms in eq.n~\eqref{Hbos}, by including the so called \textquoteleft forward scattering\textquoteright$\:$terms, i.e. the mean field interaction among the excited (real) bosons. This yields the Superstable Bogoliubov Hamiltonian:

\begin{align}
H_{SSB}&=\overbrace{\frac{\widehat{u}(0)}{2}\left[\widetilde{N}^2-\widetilde{N}\right]}^{H_{0}}+\nonumber\\
\nonumber\\
&+\sum_{\mathbf{k}\ne0}\left[\mathcal{T}(k)+\widetilde{N}_{in}\:\widehat{u}(k)\right]b^\dagger_{\mathbf{k}}b_{\mathbf{k}}+\nonumber\\
\nonumber\\
&+\frac{1}{2}\sum_{\mathbf{k}\ne0}\widehat{u}(k)\Big[b^\dagger_{\mathbf{k}}b^\dagger_{-\mathbf{k}}(\:b_{\mathbf{0}}\:)^2+b_{\mathbf{k}}b_{-\mathbf{k}}(\:b^\dagger_{\mathbf{0}}\:)^2\Big]\:.\label{HSSB}
\end{align}
\\
The interplay between the chemical potential $\mu$ and the \emph{operator} $H_0$ is shown to have important consequences in the grand canonic case \cite{AB, AB2}. In the canonic case, instead, $H_0$ behaves like a constant and, therefore, has no special relevance. In short, the results of the present work can be applied to the \emph{canonic} Superstable Bogoliubov Hamiltonian as well.

\section{Conclusions}
\label{Concl}

The drastic use of the TL, adopted by Bogoliubov in his theory, is that the operators $(b_\mathbf{0}^\dagger)^2$ and $(b_\mathbf{0})^2$, creating/annihilating pairs of (real) bosons in the free-particle ground state $|\:\mathbf{0}\:\rangle$, can be treated as c-numbers (Second Bogoliubov Approximation: SBA). This would lead one to call \emph{identical}, Fock states that are \emph{orthogonal}, like $|\:N_{in},\:N_{out}\:\rangle$ and $|\:N_{in}\pm2,\:N_{out}\:\rangle$ (Bogoliubov Canonical Approximation: BCA). Since orthogonality is a geometric property, independent from the TL, SBA and BCA look suspicious, especially because there is no \textquoteleft smallness parameter\textquoteright$\:$ controlling the quantitative aspects of the approximation. However, there are some rigorous results supporting the possibility that SBA and BCA are correct, in some sense. For instance, it has been shown \cite{G} that the pressure resulting from $H_{SBA}$, at a certain temperature, equals the one resulting from $H_{FBA}$, in the TL. Furthermore, in spite of its lack of rigor, Bogoliubov's approach yields an elegant picture of superfluidity, based on massless pseudobosons, whose theoretical developments (in particular GPT \cite{DGPS}) have been successfully tested in accurate experiments \cite{SetAl, UetAl}. As a consequence, in the current literature, the validity of SBA and BCA is accepted without special warnings \cite{AB, AB2, G, HP}, and the criticisms mostly refer to what we called the First Bogoliubov Approximation (FBA), i.e. the truncation of the interaction terms in the first-principle Hamiltonian eq.n~\eqref{Hbos} \cite{AB, AB2}.

In the present work, we have approached the problem of the exact diagonalization of $\widetilde{h}_c$ (eq.n~\eqref{Hc}). A special class of eigenstates $|\:S,\:\mathbf{k}\:\rangle_{c}$, denoted as s-eigenstates, and the corresponding eigenvalues $\mathcal{E}_S(k)$, have been expressed in an analytical form. The s-eigenstates (or s-pseudobosons) contain equal populations $n_\mathbf{k}=n_{-\mathbf{k}}$ of real bosons in the free particle states $|\pm\mathbf{k}\:\rangle$. This symmetry yields a \emph{vanishing} total momentum, which excludes the s-pseudobosons from any emission process, underlying Landau's theory of dissipation. Such possibility, instead, is accessible to different exact eigenstates, resulting from diagonalizing $\widetilde{h}_c$ in the Fock space spanned by states with asymmetry population $\eta=|n_\mathbf{k}-n_{-\mathbf{k}}|$ and total momentum $\eta\hbar\mathbf{k}$. Such non trivial diagonalization is still in progress. Preliminary calculations indicate that the resulting $\eta$-eigenstates $|\:S,\:\mathbf{k}\:\rangle_\eta$ and eigenvalues $\mathcal{E}_S(k,\eta)$ should depend on two integer labels $\eta$ and $S$. 

The exact results obtained in Section~\ref{DiagHc} make it possible to show what is right and what is not, with SBA and BCA, in the TL. The energies and, in general, all the quantities depending only on the number operator $B_\mathbf{k}^\dagger\:B_\mathbf{k}$ of BCA pseudobosons, are shown to be identical to the same quantities resulting from the s-eigenstates. The BCA ground state (pseudoboson vacuum) is the same too. However, the BCA pseudobosons themselves are not \textquoteleft contained\textquoteright$\:$ in the s-eigenstates $|\:S,\:\mathbf{k}\:\rangle_{c}$. Actually, $B_\mathbf{k}^\dagger$ and $B_\mathbf{k}$ project $|\:S,\:\mathbf{k}\:\rangle_{c}$ off, into a space orthogonal to all the $|\:S,\:\mathbf{k}\:\rangle_{c}$'s. Some differences are also found by comparison with the Gross-Pitaevskii theory (GPT): the quantum depletion of the condensate coincides with the exact result, while the thermal depletion differs by terms vanishing as $k=|\mathbf{k}|$.

A further difference between BCA and exact pseudobosons could emerge from Landau's theory of dissipation, which stems from the analogies between BCA pseudobosons and massless particles: a single BCA pseudoboson (eq.n~\eqref{|1,k>BCA}), satisfying the energy/momentum conservation laws, can be emitted by a body flowing in the Bogoliubov superfluid, which yields a decrease of the body's kinetic energy. This picture could be oversimplified, since the energy $\mathcal{E}_S(k,\eta)$ of the \emph{exact} $\eta$-pseudobosons, carrying a total momentum $\eta\hbar\mathbf{k}$, depends on $S$ too. This opens \emph{many} channels of energy dissipation, at fixed momentum change. The kinematics of the $\eta$-pseudobosons is thereby a promising field of investigation, that will be explored in forthcoming works.

\begin{appendices}
\numberwithin{equation}{section}
\section{Appendix}

As can be seen in ref.~\cite{AB} (eq.n (2.9)), the Hamitlonian eq.n~\eqref{Hc} can be rigorously expressed in terms of the operators $\beta_\mathbf{k}^\dagger$, $\beta_\mathbf{k}$ (eq.ns~\eqref{beta,beta*}): 

\begin{subequations}
\begin{align}
&\widetilde{h}_c=E_{in}+\sum_{\mathbf{k}\ne0}\widetilde{\epsilon}_1(k)\:\beta^\dagger_{\mathbf{k}}\beta_{\mathbf{k}}+\nonumber\\
\nonumber\\
&+\frac{1}{2}\sum_{\mathbf{k}\ne0}\widehat{u}(k)\Big[C(\widetilde{N}_{in})\:\beta^\dagger_{\mathbf{k}}\beta^\dagger_{-\mathbf{k}}+\:\beta_{\mathbf{k}}\beta_{-\mathbf{k}}C(\widetilde{N}_{in})\Big]\:,\label{Heff'}
\end{align}
\\
with\footnote{Notice that $\beta_{\mathbf{k}}\beta_{-\mathbf{k}}C(\widetilde{N}_{in})=C(\widetilde{N}_{in}-2)\:\beta_{\mathbf{k}}\beta_{-\mathbf{k}}$.}:

\begin{equation}
C(\widetilde{N}_{in})=\left[(\widetilde{N}_{in}+1)(\widetilde{N}_{in}+2)\right]^{1/2}\label{C}\:,
\end{equation}
\end{subequations}
\\
From eq.ns~\eqref{Heff'}, one gets the matrix elements:

\begin{align}
&\langle\:N_{out},\:\:N_{in}\:|\widetilde{h}_c |\:N_{in}'\:,\:N_{out}'\:\rangle=\nonumber\\
\nonumber\\
&=\delta_{\{n_\mathbf{k}\},\:\{n_\mathbf{k}'\}}\delta_{N_{in},\:N_{in}'}\Big[E_{in}+\sum_{\mathbf{k}\ne0}\left[\mathcal{T}(k)+N_{in}\:\widehat{u}(k)\right]n_\mathbf{k}\Big]+\nonumber\\
\nonumber\\
&+\frac{1}{2}\sum_{\mathbf{k}\ne0}\widehat{u}(k)\Big[C(N_{in})\underbrace{\langle\:N_{out},\:\:N_{in}\:|\beta^\dagger_{\mathbf{k}}\beta^\dagger_{-\mathbf{k}}|\:N_{in}'\:,\:N_{out}'\:\rangle}_{\propto\:\delta_{N_{in}',\:N_{in}+2}}+\nonumber\\
\nonumber\\
&+C(N_{in}')\underbrace{\langle\:N_{out},\:\:N_{in}\:|\beta_{\mathbf{k}}\beta_{-\mathbf{k}}|\:N_{in}'\:,\:N_{out}'\:\rangle}_{\propto\:\delta_{N_{in}',\:N_{in}-2}}
\Big]\label{MatrixEl}\:,
\end{align}
\\
between states of the Fock space of interest. At this stage, the Bogoliubov Canonical Approximation (BCA) proceeds in two steps: first, one sets $|\:N_{in}\pm2,\:N_{out}\:\rangle=|\:N_{in},\:N_{out}\:\rangle$, thanks to which it is possible to treat the operator $\widetilde{N}_{in}$ as a c-number; second, one sets $N_{in}=N$, which is a less serious, zero-order approximation in $\alpha$. Under those assumptions, the \emph{canonic} Hamiltonian resulting from eq.n~\eqref{MatrixEl} is $H_{BCA}$ eq.n~\eqref{HBCA}, once noticed that $C(N_{in})\rightarrow N_{in}$ in the TL (eq.n~\eqref{C}).

\section{Appendix}

In the following calculations, the dependence on $\mathbf{k}$, $T$ of $S$ and $x$ will be omitted for brevity. Making use of the normalization condition, one can write, from eq.n~\eqref{< n(k) >T}:

\begin{subequations}
\begin{align}
&\langle\:n_{\mathbf{k}}\:\rangle_T=\frac{\sum_{j=0}^\infty x^{2j}\:j\left[\sum_{s=0}^SC_s\:j^s\right]^2}{\sum_{j=0}^\infty x^{2j}\left[\sum_{s=0}^SC_s\:j^s\right]^2}=\nonumber\\
\nonumber\\
&=\frac{\sum_{s,s'=0}^SC_sC_{s'}\sum_{j=0}^\infty x^{2j}j^{s+s'+1}}{\sum_{s,s'=0}^SC_sC_{s'}\sum_{j=0}^\infty x^{2j}j^{s+s'}}=\nonumber\\
\nonumber\\
&=\frac{\sum_{s,s'=0}^SC_sC_{s'}\sum_{j=0}^\infty x^{2j}(j+1)^{s+s'+1}}{\sum_{s,s'=0}^SC_sC_{s'}\sum_{j=0}^\infty x^{2j}(j+1)^{s+s'}+C_0^2/(x^2-x^4)}=\label{< n(k) >T2}\\
\nonumber\\
&=\frac{\sum_{s,s'=0}^SC_sC_{s'}\Phi\left(x^2,\:-(s+s'+1),\:1\right)}{\sum_{s,s'=0}^SC_sC_{s'}\Phi\left(x^2,\:-(s+s'),\:1\right)+C_0^2/(x^2-x^4)}=\label{< n(k) >T3}\\
\nonumber\\
&=\mathcal{R}\left(x^2,\:2S+1,\:2S\right)\times\nonumber\\
\nonumber\\
&\times\frac{1+C^{-2}_S\sum_{s+s'=0}^{2S-1}C_sC_{s'}\mathcal{R}\left(x^2,\:s+s'+1,\:2S+1\right)}{1+C^{-2}_S\left[\sum_{s+s'=0}^{2S-1}C_sC_{s'}\mathcal{R}\left(x^2,\:s+s',\:2S\right)-C_0^2\mathcal{R}_0(x^2,\;2S)\right]}\:,\label{< n(k) >T4}
\end{align}
\end{subequations}
\\
where the definition~\eqref{Phi} has been used, in passing from eq.n~\eqref{< n(k) >T2} to  eq.n~\eqref{< n(k) >T3}, and:

\begin{subequations}
\label{Rstrange}
\begin{align}
\mathcal{R}\left(x^2,\:M,\:L\right)&=\frac{\Phi\left(x^2,\:-M,\:1\right)}{\Phi\left(x^2,\:-L,\:1\right)}=\nonumber\\
\nonumber\\
&=\frac{M!}{L!}\left(\frac{x^2}{1-x^2}\right)^{M-L}\left[1+\mathrm{o}(1-x^2)\right]\\
\nonumber\\
\mathcal{R}_0(x^2,\;2S)&=\frac{1}{(x^2-x^4)\Phi\left(x^2,\:-2S,\:1\right)}=\nonumber\\
\nonumber\\
&=\frac{(1-x^2)^{2S+1}}{(2S)!x^{2(2S+1)}}\left[1+\mathrm{o}(1-x^2)\right]\:,
\end{align}
\end{subequations}
\\
according to eq.ns~\eqref{H-L-Phi}. On applying eq.ns~\eqref{Rstrange} to eq.n~\eqref{< n(k) >T4}, it is easily seen that:

\begin{equation*}
\label{< n(k) >T5}
\langle\:n_{\mathbf{k}}\:\rangle_T=\frac{2\:S(T,\:k)+1}{1-x^2(k)}\left[1+\mathrm{o}(1-x^2)\right]\:,
\end{equation*}
\\
which yields eq.n~\eqref{< n(k) >T'}.
\end{appendices}

\end{document}


\title{Errata Corrige to \textquoteleft Exact canonic eigenstates of the truncated Bogoliubov Hamiltonian in an interacting bosons gas\textquoteleft}
\author{Loris Ferrari \\ Department of Physics and Astronomy of the University (DIFA) \\Via Irnerio, 46,40127, Bologna, Italy}
\maketitle
e-mail: loris.ferrari@unibo.it
telephone: ++39-051-2095109\newline

In the author's paper \textquoteleft Exact canonic eigenstates of the truncated Bogoliubov Hamiltonian in an interacting bosons gas\textquoteleft$\:$(Physica B, 2016, pg 38-44), an error occurred in identifying the eigenvalue $\mathcal{E}_S(k)$ (eq.n (25c)) of the Hamiltonian $\widetilde{h}_c(\mathbf{k})$ (eq.n (17b)) with the eigenvalue $E_S(k)$ of Hamiltonian $H_c$ (eq.n (5)). The correct claim is, instead, $E_S(k)=2\mathcal{E}_S(k)$, since in the sum eq.n (17a) the eigenvalue of $\widetilde{h}_c(\mathbf{k})=\widetilde{h}_c(-\mathbf{k})$ is to be counted twice. Due to this error, the symbol $\mathcal{E}_S(k)$ in the 15-th line (from bottom) of pg 40 must be replaced by $E_S(k)$, and the last sentence in pg 40: /\emph{In Section 4 it will be shown that ... of the s-eigenstates.}/ must be modified as follows:\newline
\\
\emph{In Section 4 it will be shown that $E^{BCA}_S(k)=E_S(k)/2$, i.e. that the BCA energy eigenvalues are \emph{half} those of the s-eigenstates.}\newline
\\
In pg 41, just after eq.n (25c) and the sentence /\emph{with $k$ and $N$ restored.}/, one should insert: \newline
\\
\emph{Notice that:}

\begin{equation*}
E_S(k)=2\mathcal{E}_S(k), \quad \quad  (25d)
\end{equation*}
\\
\emph{i.e. the eigenvalues $E_S(k)$ of $H_c$ are twice as large as those of $\widetilde{h}_c$, since they must be counted twice in the sum eq.n (17a).}\newline
\\
In pg 42, just after \textbf{4. Comparison and discussion}, the sentence: /\emph{the \emph{exact} eigenvalues ... (Eq. (6)).}/ must be replaced by:\newline 
\\
\emph{the \emph{exact} eigenvalues $E_S(k)$, corresponding to the s-eigenstates $|\:S,\:\mathbf{k}\:\rangle_c$, are twice as large as the energies $E_S^{BCA}(k)$ obtained in Section 2 (Eq. (6)).}\newline
\\
At the end of pg 42 (7-th line from bottom), the whole initial sentence /\emph{In short: ... are \emph{exact}.}/ must be dropped.\newline
\\
Eq.n (35) must be corrected by replacing $\epsilon(k)$ with $2\epsilon(k)$ in the exponential.\\

The corrections indicated do not change the meaning and the spirit of the paper. Rather, they support even more the main result that the exact eigenstates of $H_c$ and those obtained under BCA are quite different, since at this stage it is clear that the eigenvalues too are different ($E_S(k)=2E_S^{BCA}(k)$). This is expected to lead to non trivial physical consequences, as will be shown in a forthcoming paper.